# Evaluating the Effectiveness of Pre-Trained Audio Embeddings for Classification of Parkinson's Disease Speech Data

*Emmy Postma, Cristian Tejedor-Garcia*

Centre for Language Studies, Radboud University Nijmegen, the Netherlands
emmy.postma@ru.nl, cristian.tejedorgarcia@ru.nl

## Abstract

Speech impairments are prevalent biomarkers for Parkinson's Disease (PD), motivating the development of diagnostic techniques using speech data for clinical applications. Although deep acoustic features have shown promise for PD classification, their effectiveness often varies due to individual speaker differences, a factor that has not been thoroughly explored in the existing literature. This study investigates the effectiveness of three pre-trained audio embeddings (OpenL3, VGGish and Wav2Vec2.0 models) for PD classification. Using the NeuroVoz dataset, OpenL3 outperforms others in diadochokinesis (DDK) and listen and repeat (LR) tasks, capturing critical acoustic features for PD detection. Only Wav2Vec2.0 shows significant gender bias, achieving more favorable results for male speakers, in DDK tasks. The misclassified cases reveal challenges with atypical speech patterns, highlighting the need for improved feature extraction and model robustness in PD detection.

## 1. Introduction

Recent research has established speech as a critical and cost-effective biomarker of Parkinson's disease (PD) [1, 2, 3]. Given that speech impairments are common symptoms of PD and often emerge early in the disease progression, researchers have increasingly utilized speech data in machine learning (ML) approaches to develop models capable of classifying PD from speech recordings [4]. Various techniques have been explored for extracting features from speech data to support these classification tasks. These range from conventional features, such as jitter, shimmer and MFCCs, to deep acoustic features derived from deep learning (DL) models, a process known as Deep Acoustic Feature Extraction (DAFE) [2].

Recent studies highlight the value of deep features for PD classification when combined with open-source audio models and ML classifiers, some of which primarily capture acoustic features, while others more effectively extract speech-related information. Although direct comparisons are challenging due to variations in datasets and evaluation methods, these approaches consistently outperform random classification and, in some cases, achieve an estimated clinical diagnostic accuracy (ACC) of 0.90–0.97 [5]. For instance, Syed et al. [6] achieved an average ACC=0.79 in six versions of a diadochokinesis (DDK) task by using the pre-trained OpenL3 (OL3) model for feature extraction. In the same study, Syed et al. reported an average ACC=0.71 using VGGish (VGG) embeddings derived from speech recordings collected during a sentence-reading task. The VGG model with a Support Vector Machine (SVM) classifier was also used in [7], achieving ACC=0.95 in voice disorder classification. Kurada and Kurada [8], achieved ACC=0.83 for PD classification with the same model embeddings and an Extremely Randomized Trees (ERT) classifier. Favaro et al. [9] reported AUCs of 0.57 to 1.00 using Wav2Vec2.0 (W2V2) embeddings. Also using W2V2, Klempíř et al. [10] found AUCs of 0.69 to 0.98. Alshammri et al. [11] reported up to ACC=0.88 for PD classification using K-Nearest Neighbors (KNN) with voice features.

While these results highlight the potential of deep features for PD classification, their performance may still be constrained by several challenges, including the substantial variability of speech impairments between individuals. Although PD is widely recognized for causing such speech impairments, they differ significantly between speakers, underscoring the need for both individual and gender-specific considerations in PD research [12, 2]. For instance, changes in intonation and speech rate are more relevant for diagnosing PD in females, while an increased mean fundamental (or pitch) frequency (i.e., F0) is often an early sign in males [12]. Botelho et al. [13] further identified gender-specific differences in acoustic features, with F0 features being more significant for males and harmonics-to-noise ratio, jitter and shimmer features more relevant for females. Their study reported a PD detection ACC=0.67 for females compared to 0.75 for males. Furthermore, males have approximately double the risk of developing PD [14], leading to less available data from female patients, who also tend to develop PD about two years later than males [15]. Finally, the speech features most discriminative for PD vary between genders, and using mixed-gender data may reduce model performance, emphasizing the need for gender-specific approaches to improve the reliability and accuracy of PD classification models. Jeancolas et al. [16] found that, using MFCCs, PD detection was poorer in females (Equal Error Rate (EER): 40-45%) than males (22-36%). And also when using x-vectors, performance for females was worse (EER: 30-39%), compared to males (22%). This suggests that features effective for males may not be as effective for females, and vice versa [17, 18].

In this work, we present two key contributions to the field of PD classification using speech data. First, we evaluate and compare the effectiveness of three widely used open-source pre-trained audio models—OL3, VGG and W2V2— which show the most promising results in extracting features from embeddings for binary PD classification. Secondly, we investigate the impact of speaker differences on classification performance, providing insight into how speaker-level variability influences model outcomes. Therefore, in this study, we will explore the following two research questions (*RQs*):

*RQ1*. How do pre-trained OL3, VGG and W2V2 embeddings compare in terms of binary PD classification performance using DDK (*RQ1a*) and LR (*RQ1b*) speech tasks?

*RQ2*. How does the performance of OL3, VGG and W2V2 embeddings vary among speakers in the classification of PD?

## 2. Methodology

### 2.1. Dataset

We used the publicly available NeuroVoz[1] dataset [19], which includes voice recordings and PD-related metadata from 108 Castilian Spanish speakers—53 with PD and 55 healthy controls (HC). Metadata includes gender, age, vocal and cephalic tremors, mandibular tremor, sialorrhea, dysphagia, hypophonic voice, time since diagnosis (TAD) and unified PD rating scale (UPDRS) score. The PD group (mean age: $71.1 \pm 10.6$) had an official diagnosis and was on medication during data collection, with an average TAD of 6.9 years. The HC group (mean age: $64.0 \pm 10.3$) had no neurological disorders. We used data from two speech tasks: The /pa-ta-ka/ diadochokinetic (DDK) test and Listen-and-Repeat (LR) exercises. In the DDK task, participants repeated "pa-ta-ka" for at least five seconds, with mean durations of $11.11 \pm 4.64$ seconds (PD) and $13.06 \pm 5.46$ seconds (HC). In the LR task, participants repeated 16 predefined phrases, with mean phrase durations of $3.26 \pm 2.45$ seconds (PD) and $3.70 \pm 2.74$ seconds (HC). The recordings were made at a 44.1 kHz sampling rate [19].

### 2.2. Experimental Procedure

Figure 1 illustrates the experimental workflow. Embeddings from the NeuroVoz dataset were extracted separately from DDK and LR recordings using pre-trained OL3, VGG, and W2V2 models. These models produced window-based embeddings, yielding multiple embeddings per recording. To derive a single representation per recording, the embeddings were summarized using average pooling, specifically by computing the mean across all windows, as recommended by Wilkinghoff [20]. The resulting embeddings were used to train and evaluate three classifiers per model, SVM, KNN, and ERT (nine pipelines in total), consistent with the current literature on speech-based PD classification [7, 8, 11]. A speaker-independent 5-fold nested cross-validation (nested-CV) approach was adopted per pipeline of model-classifier pair, ensuring that embeddings from the same participant remained in the same fold for both training and testing, without overlap.[2] To investigate differences between speakers in classification performance, all the nine pipelines were separately trained and evaluated on features derived from male and female voices, followed by a manual inspection of individual cases. Performance was assessed using accuracy (ACC) and the area under the receiver operating characteristic curve (AUC) to measure prediction agreement. The Matthews correlation coefficient (MCC) was also used as a separate metric to evaluate the prediction agreement between pipelines.

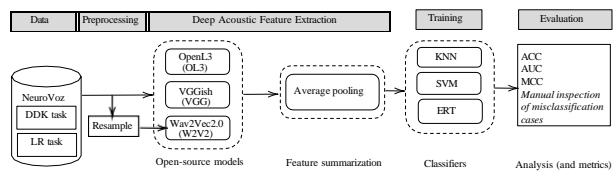

Figure 1: *Steps of the experimental procedure.*

---

[1] https://zenodo.org/record/13647600 (last visited: 2025-02-01)
[2] https://gitlab.science.ru.nl/epostma/thesis (last visited: 2025-02-01)

### 2.3. Pre-trained Audio Models

In this work, we used three widely used open-source pre-trained audio models, which have demonstrated the best performance in similar classification tasks in previous studies [9, 21, 22], but have not been directly compared on the same PD classification task and dataset yet. The first one, OL3, is a pre-trained variant of the L3-Net model, trained through self-supervised learning on the AudioSet[3] dataset. It resamples raw audio to 48 kHz and, in this study, the default version—trained on musical sounds—was used. This model generates 6144-dimensional embeddings per window, extracted from overlapping 1-second windows with a 0.1-second hop size. Second, we employed the VGG model, pre-trained on AudioSet audio clips using supervised learning. It processes Mel spectrograms through five convolutional layers followed by three dense layers [23]. Input audio was resampled to 16 kHz and embeddings were extracted using a default hop size of 0.96 seconds. The third model, W2V2, is a self-supervised model designed for automatic speech recognition (ASR). The large variant consists of seven convolutional layers and 24 transformer layers.[4] W2V2 requires input resampling to 16 kHz. We used the pre-trained wav2vec2-large-xlsr-53-spanish[5] model by Meta from HuggingFace, which was trained on 56,000 hours of multilingual, unlabeled data from datasets such as Multilingual LibriSpeech [24], CommonVoice, [25] and BABEL [26].

## 3. Results and Discussion

### 3.1. Models Performance

To answer *RQ1*, we compared the performances of all three embedding model (OL3, VGG, W2V2) and classifier (SVM, KNN, ERT) combinations on both DDK and LR data. Table 1 presents the mean and standard deviation (SD) of the ACC and AUC scores across the five outer folds of the nested-CV. To evaluate potential violations of normality, Shapiro-Wilk and Kolmogorov-Smirnov tests were performed. The results did not indicate significant deviations from normality, as all distributions obtained p-values of 0.05 or greater. Consequently, Welch's t-tests were utilized for pairwise comparisons of the embeddings, focusing on the best-performing classifier for each model.

Table 1: *Performance scores for DDK and LR data*

| Model | Classi. | DDK | | LR | |
|---|---|---|---|---|---|
| | | ACC ± SD | AUC ± SD | ACC ± SD | AUC ± SD |
| OL3 | SVM | 0.85 ± 0.05 | 0.77 ± 0.37 | **0.82 ± 0.04** | **0.90 ± 0.03** |
| | KNN | 0.84 ± 0.11 | 0.91 ± 0.08 | 0.76 ± 0.03 | 0.81 ± 0.03 |
| | ERT | **0.85 ± 0.05** | **0.97 ± 0.08** | 0.71 ± 0.02 | 0.80 ± 0.02 |
| VGGish | SVM | **0.81 ± 0.10** | 0.86 ± 0.05 | **0.78 ± 0.05** | **0.85 ± 0.05** |
| | KNN | 0.75 ± 0.09 | 0.81 ± 0.04 | 0.72 ± 0.03 | 0.77 ± 0.04 |
| | ERT | 0.78 ± 0.08 | 0.86 ± 0.06 | 0.73 ± 0.03 | 0.79 ± 0.04 |
| W2V2 | SVM | **0.71 ± 0.08** | 0.65 ± 0.21 | **0.76 ± 0.04** | **0.84 ± 0.02** |
| | KNN | 0.67 ± 0.07 | 0.70 ± 0.08 | 0.74 ± 0.02 | 0.78 ± 0.03 |
| | ERT | 0.63 ± 0.11 | 0.63 ± 0.07 | 0.70 ± 0.02 | 0.77 ± 0.03 |

Table 1 shows that using DDK recordings (*RQ1a*), the best-performing classifiers in terms of ACC for each model were OL3-SVM and OL3-ERT (both ACC=0.85), VGG-SVM (ACC=0.81) and W2V2-SVM (ACC=0.71). Pairwise comparisons, using a Welch's t-test showed statistically significant differences in ACC between OL3 and W2V2 (T=2.86; df=6.93;

---

[3] https://github.com/marl/openl3 (last visited: 2025-02-01)
[4] https://jonathanbgn.com/2021/09/30/illustrated-wav2vec-2.html (last visited: 2025-02-01)
[5] https://huggingface.co/facebook/wav2vec2-large-xlsr-53-spanish (last visited: 2025-02-01)

p=0.025); whereas no significant differences were found between OL3 and VGG (T=0.71; df=6.36; p=0.50) and between VGG and W2V2 (T=1.62; df=7.84; p=0.15). Considering AUC, the best-performing classifiers were OL3-ERT (AUC=0.97), VGG-SVM (AUC=0.86) and W2V2-KNN (AUC=0.70). Pairwise comparisons showed that OL3 scored significantly higher than VGG (T=3.33; df=5.75; p=0.017) and VGG scored significantly higher than W2V2 (T=3.48; df=7.18; p=0.0098). Furthermore, OL3 scored significantly higher than W2V2 (T=6.50; df=4.90; p=0.0014). These results indicate that OL3 extracts the most effective embeddings for PD classification using DDK speech data, outperforming VGG and W2V2. While the superiority of OL3 over VGG is consistent with prior research findings [6, 27], to the best of our knowledge, no existing studies have conducted a performance comparison for PD classification using W2V2 embeddings alongside these models in the literature. The ACC values for OL3 (0.85) and VGG (0.71) in this study exceed those reported in [6] (0.73 and 0.70, respectively), with a slightly smaller performance gap. These differences may arise from dataset variations (PC-GITA vs. NeuroVoz) or classifier choice (SVM vs. Logistic Regression). For W2V2 in a similar DDK task, Klempíř et al. [10] reported ACC=0.61 and AUC=0.69, closely matching our results (ACC: 0.63–0.71, AUC: 0.63–0.72).

Table 1 also shows that using LR data (*RQ1b*), the best-performing classifiers in terms of ACC for each model were OL3-SVM (ACC=0.82), VGG-SVM (ACC=0.78) and W2V2-SVM (ACC=0.76). Pairwise comparisons using Welch's t-tests showed one single statistically significant difference in ACC between models (OL3 vs. W2V2, T=2.41; df=6.64; p=0.049). Considering AUC, the best-performing classifiers were OL3-SVM (AUC=0.90), VGG-SVM (AUC=0.85) and W2V2-SVM (AUC=0.84). T-tests showed only one case of a statistically significant difference (as for ACC) between OL3 and W2V2 (T=2.67; df=7.26; p=0.031).

For both DDK and LR data, OL3 embeddings outperformed W2V2, despite W2V2 being trained on speech data specifically. This could be due to W2V2's emphasis on lexical information for ASR tasks [9], while OL3, trained on musical data, better captures non-linguistic features such as timbre, rhythm and sound textures. Given that PD-related acoustic changes are more linked to sound production than linguistic content [28, 29], OL3 may be more effective, as all participants repeated the same words and sentences. Our results for LR embeddings are consistent with those of Syed et al. [6] for short utterances task: With OL3 embeddings, Syed et al. found ACC between 0.71–0.77, and VGG embeddings resulted in an average ACC=0.71. These results are slightly lower than the 0.82 and 0.78 ACC scores we found for OL3 and VGG, respectively. Using text-reading recordings and W2V2 embeddings, Klempíř et al. [10] reported a significantly higher ACC=0.95 and AUC=0.98, compared to our results (ACC: 0.69–0.76, AUC: 0.76–0.84). These differences may stem from their use of an English-only W2V2 model, a different dataset, and a different classifier.

Figure 2 shows the pairwise correlations between the pipelines, measured using the MCC metric to evaluate how well the models agreed on their predictions for individual recordings. The highest agreement using DDK data (left matrix, Figure 2) was between OL3-SVM and OL3-KNN (MCC=0.78), while the lowest was between VGG-ERT and W2V2-ERT (MCC=0.21). Overall, classifiers using VGG and W2V2 embeddings showed the least agreement, with correlations generally higher within the same DL model. These results suggest that classifiers trained with the same DDK embeddings (*RQ1a*) tend to produce more similar predictions than identical classifiers trained on different embeddings. Furthermore, although OL3 and VGG both score relatively high on ACC (Table 1), their agreement is only fair (between 0.47 and 0.6). This indicates that OL3 may be more effective in some cases, such as recordings that are more challenging to classify, while VGG may perform better with others.

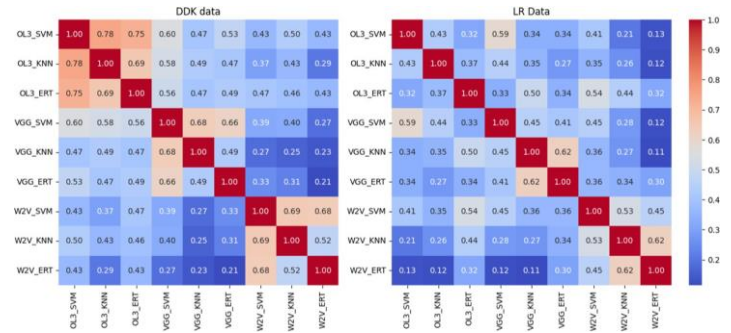

Figure 2: *Agreement (MCC scores) between models trained on DDK data (left) and LR data (right).*

The matrix on the right side of Figure 2 shows that for LR data (*RQ1b*) the highest agreement was found between W2V2-ERT and W2V2-KNN (MCC=0.62) and between VGG-ERT and VGG-KNN (MCC=0.62), while the lowest was between VGG-KNN and W2V2-ERT (MCC=0.11). Overall, agreement was lower between classifiers using OL3 or VGG embeddings and those using W2V2 embeddings. The pattern that was found for the DDK data, with higher MCC scores between classifiers of the same model, seems to be absent with LR data. Again, the best-performing models in accuracy (OL3 and VGG) show a (maximum) moderate agreement of MCC=0.59; suggesting they perform best in different cases.

### 3.2. Differences Between Speakers

To answer *RQ2*, we analyzed the performance of the same three classifiers when trained on embeddings generated from male and female voices separately. Furthermore, we inspected how often a participant was misclassified and compared the most problematic cases to the most successful ones.

Figure 3 presents the AUC scores for all DDK pipelines trained on three cases: all data, male only and female only data, separately. To test whether the differences in performance between classifiers trained on isolated genders are statistically significant, Welch's t-tests were performed on the AUC scores for every pipeline. W2V2-SVM was the only pipeline for which a statistically significant difference was found. For this model, the performance of the classifier trained on embeddings generated from male voices (AUC=0.82; SD=0.09), was significantly higher (T=3.76; df=5.64; p=0.01) than the performance on embeddings taken from female voices (AUC=0.45; SD=0.20). The W2V2-SVM pipeline result shows a clear bias toward male speakers, consistent with Fraile et al. [30], who found that using MFCCs instead of DAFE made speech pathology detection more effective for males than females. This may be due to the greater variability in women's voices [30], making abnormalities harder to detect. Furthermore, since W2V2 is specifically trained on speech data, an imbalanced gender distribution in the training data could be a contributing factor. For instance, if W2V2 were predominantly trained on male voices, this might limit its ability to learn accurate audio representations for both genders. While the Multilingual LibriSpeech dataset [24] is gender-balanced, the gender distributions for the BABEL [26] and CommonVoice [25] datasets remain unknown.

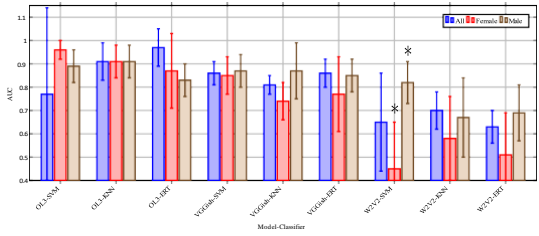

Figure 3: *Mean AUC values and standard deviations for DDK Data. For each set of bars, the first bar represents All, the second Female and the third Male. Statistically significant differences are indicated by the * symbol.*

Figure 4 displays the AUC scores for all pipelines trained on LR data, including the full dataset, male-only data and female-only data. T-tests indicated no significant performance differences between models trained on embeddings from male and female voices. These findings suggest that LR speech task data results in comparable model performance across genders.

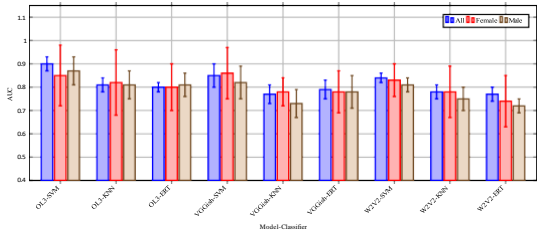

Figure 4: *Mean AUC values and standard deviations for LR Data. For each set of bars, the first bar represents All, the second Female and the third Male.*

In addition to the effect of gender on the performance of our models, we also explored possible other factors that may cause variation in performances between individual speakers. To determine whether any recordings were misclassified by all pipelines, a heatmap was created to display the misclassification cases from the DDK recordings. Figure 5 shows that the DDK recording from participant 34 (HC group) was misclassified by eight pipelines, suggesting that it was the most challenging HC case to classify correctly. In contrast, eighteen participants were correctly classified by all nine pipelines.

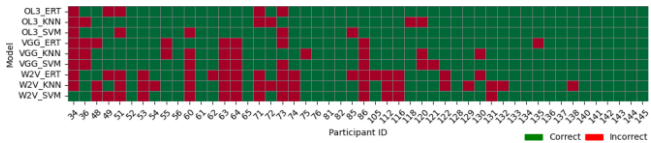

Figure 5: *Misclassification cases in HC group using DDK data.*

Figure 6 shows that the DDK recording from participant 111 (PD group) was misclassified by all nine pipelines. This case seemed to be the hardest PD speaker to classify. Oppositely, twelve participants were correctly classified by all pipelines.

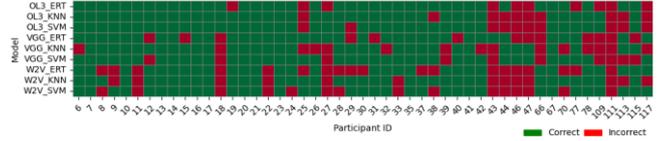

Figure 6: *Misclassification cases in PD group using DDK data.*

In an effort to understand why some speech recordings were frequently misclassified, we examined the five most problematic cases more closely (participants 111, 43 and 47 in Figure 5, and participants 34 and 73 in Figure 6) and compared them to five consistently correctly classified cases (participants 52, 56 in Figure 5 and participants 7, 13 and 14 in Figure 6). Hypophonia, or reduced voice volume, was more common among the misclassified PD recordings and was not present in the easy group. However, there were no obvious differences in age, gender, TAD, or clinical severity (measured by UPDRS scores) between the two groups.

Listening to the recordings revealed distinct speech patterns among participants. Hard-case speakers from the HC group (e.g., participants 34 and 73 in Figure 5) spoke slowly and less fluently, often pausing or struggling, unlike easy-case speakers also from HC (e.g., 52 and 56 in Figure 5), who spoke fluently. In the PD group, hard-case participant 111 spoke softly but quickly and without errors, contrasting with easy-case PD speakers like 7 (few errors) and 13 (slow pace). Similarly, participant 43 (hard-case; PD) spoke faster than 13, made fewer errors than 7, and was louder than 14. Participant 47 (hard-case; PD) had a similar pace and volume to 14 but made occasional errors. Challenging cases often exhibited atypical traits, such as soft voices and stuttering in hard-case HC speakers or smooth pronunciation in hard-case PD speakers. These characteristics may contribute to misclassification; however, further research involving clinical expertise is required to confirm this.

## 4. Conclusions

The findings of this study highlight the overall superior performance of OL3 pre-trained audio embeddings for PD classification using speech data from DDK and LR tasks from the NeuroVoz dataset. Furthermore, despite OL3 and VGG achieving a high ACC, their low MCC agreement indicates that there is a clear speaker variance in PD data due to the complexity of PD speech patterns. For LR data, the highest MCC agreement occurred among classifiers using the same embeddings, while cross-embedding agreement remained low. This emphasizes the unique strengths of each model and the critical role of correct selection of embedding classifier.

Our work also revealed a statistically significant gender bias in the W2V2-SVM pipeline with DDK data, clearly yielding better results for male speakers than female speakers. However, models performed equally well between genders for the LR task. Analysis of individual misclassified cases identified hypophonia and other speech characteristics (e.g., fluency, pace) as potential contributors to classification challenges. These findings underscore the need for further research to refine feature extraction and model training, particularly for "hard" cases of individual speakers, to improve the robustness and fairness of realistic PD classification systems in clinical settings. Future work should focus on integrating diverse datasets and investigating advanced embedding combinations to improve model generalizability and performance across varied speech patterns.

## 5. Acknowledgements

This publication is part of the project Responsible AI for Voice Diagnostics (RAIVD) with file number NGF.1607.22.013 of the research programme NGF AiNed Fellowship Grants which is financed by the Dutch Research Council (NWO). The authors of this work would also like to thank Prof. Dr. Martha Larson for her insightful feedback regarding the individual differences analysis.